\documentclass[iop]{emulateapj}
\usepackage{amsmath}
\usepackage{graphicx}
\usepackage{epsfig}

\begin{document}


\title{GNOSIS: THE FIRST INSTRUMENT TO USE FIBER BRAGG GRATINGS FOR OH SUPPRESSION}
\author{Christopher Q. Trinh\altaffilmark{1}, Simon C. Ellis\altaffilmark{2,1}, Joss Bland-Hawthorn\altaffilmark{1,3}, Jon S. Lawrence\altaffilmark{2,4}, Anthony J. Horton\altaffilmark{2}, Sergio G. Leon-Saval\altaffilmark{3}, Keith Shortridge\altaffilmark{2}, Julia Bryant\altaffilmark{1}, Scott Case\altaffilmark{2}, Matthew Colless\altaffilmark{2}, Warrick Couch\altaffilmark{5}, Kenneth Freeman\altaffilmark{6}, Hans-Gerd L\"{o}hmannsr\"{o}ben\altaffilmark{7}, Luke Gers\altaffilmark{2}, Karl Glazebrook\altaffilmark{5}, Roger Haynes\altaffilmark{8}, Steve Lee\altaffilmark{2}, John O'Byrne\altaffilmark{1}, Stan Miziarski\altaffilmark{2}, Martin M. Roth\altaffilmark{8}, Brian Schmidt\altaffilmark{6}, Christopher G. Tinney\altaffilmark{9}, Jessica Zheng\altaffilmark{2}}
\altaffiltext{1}{Sydney Institute for Astronomy, School of Physics, The University of Sydney, NSW 2006, Australia; c.trinh@physics.usyd.edu.au}
\altaffiltext{2}{Australian Astronomical Observatory, 105 Delhi Road, North Ryde, P.O. Box 915, NSW 1670, Australia}
\altaffiltext{3}{Institute of Photonics and Optical Science, School of Physics, The University of Sydney, NSW 2006, Australia}
\altaffiltext{4}{Department of Physics and Astronomy, Macquarie University, NSW 2109}
\altaffiltext{5}{Centre for Astrophysics and Supercomputing, Swinburne University of Technology, P.O. Box 218, Hawthorn, VIC 3122, Australia}
\altaffiltext{6}{Research School of Astronomy and Astrophysics, Australian National University, Weston Creek, ACT 2611, Australia}
\altaffiltext{7}{innoFSPEC-Institut f\"{u}r Chemie/Physikalische Chemie, Universit\"{a}t Potsdam, Karl-Liebknecht-Strasse 24-25, D-14476 Potsdam-Golm, Germany}
\altaffiltext{8}{innoFSPEC-Leibniz-Institut f\"{u}r Astrophysik Potsdam, An der Sternwarte 16, D-14482 Potsdam, Germany}
\altaffiltext{9}{Department of Astrophysics, School of Physics, University of New South Wales, NSW 2052, Australia}

\shorttitle{Fiber Bragg gratings for OH suppression}
\shortauthors{Trinh et al.}

\begin{abstract}
The near-infrared is an important part of the spectrum in astronomy, especially in cosmology because the light from objects in the early universe is redshifted to these wavelengths. However, deep near-infrared observations are extremely difficult to make from ground-based telescopes due to the bright background from the atmosphere. Nearly all of this background comes from the bright and narrow emission lines of atmospheric hydroxyl (OH) molecules. The atmospheric background cannot be easily removed from data because the brightness fluctuates unpredictably on short timescales. The sensitivity of ground-based optical astronomy far exceeds that of near-infrared astronomy because of this long-standing problem. GNOSIS is a prototype astrophotonic instrument that utilizes ``OH suppression fibers" consisting of fiber Bragg gratings and photonic lanterns to suppress the 103 brightest atmospheric emission doublets between 1.47 and 1.7\,$\micron$. GNOSIS was commissioned at the 3.9\,m Anglo-Australian Telescope with the IRIS2 spectrograph to demonstrate the potential of OH suppression fibers, but may be potentially used with any telescope and spectrograph combination. Unlike previous atmospheric suppression techniques GNOSIS suppresses the lines before dispersion and in a manner that depends purely on wavelength. We present the instrument design and report the results of laboratory and on-sky tests from commissioning. While these tests demonstrated high throughput ($\approx$ 60\%) and excellent suppression of the skylines by the OH suppression fibers, surprisingly GNOSIS produced no significant reduction in the interline background and the sensitivity of GNOSIS+IRIS2 is about the same as IRIS2. It is unclear whether the lack of reduction in the interline background is due to physical sources or systematic errors as the observations are detector noise dominated. OH suppression fibers could potentially impact ground-based astronomy at the level of adaptive optics or greater. However, until a clear reduction in the interline background and the corresponding increasing in sensitivity is demonstrated optimized OH suppression fibers paired with a fiber-fed spectrograph will at least provide a real benefit at low resolving powers. 
\end{abstract}

\keywords{atmospheric effects -- infrared: diffuse background -- instrumentation: miscellaneous}

\section{Introduction}
Near-infrared (NIR) observations (0.9--2.5\,$\micron$) are important in practically all areas of astronomy and astrophysics. For example, low-mass stars and brown dwarfs emit a substantial fraction of their light at these wavelengths and NIR spectroscopy is the most efficient way to study these objects. NIR spectroscopy is also one of the best ways to study the early universe because optical and ultraviolet emission lines from distant galaxies are redshifted to NIR wavelengths due to Hubble expansion. Unfortunately, deep NIR observations from the ground are extremely difficult to make due to the presence of a bright atmospheric background. The background in the range 0.9--1.8\,$\micron$ is predominantly from the de-excitation of atmospheric hydroxyl (OH) molecules at an altitude of $\approx$ 90\,km \citep{meinel1950,dufay1951}. The NIR background is $\approx$ 1000 times brighter than the optical background and cannot be simply subtracted from astronomical observations \citep{davies2007} because its brightness fluctuates on short timescales \citep{ramsay1992}. Solving the NIR sky background problem is an important challenge in observational astronomy.  

\begin{figure*}[!ht]
\center
\includegraphics[width=0.95\textwidth]{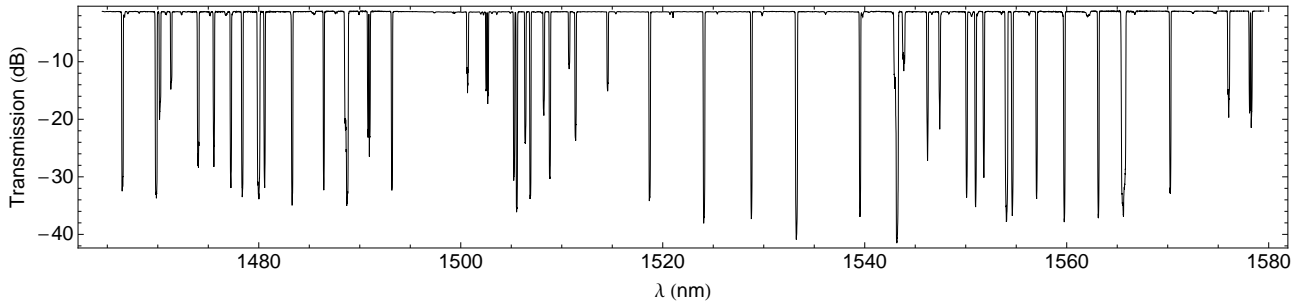} 
\caption{Wavelength response of a single aperiodic FBG with 50 notches. GNOSIS uses two aperiodic FBGs in series to suppress the 103 brightest OH doublets in the range 1.47--1.7\,$\micron$.\label{fig:afbgresponse}}
\end{figure*}

Previous attempts at a ground-based solution have not been able to suppress OH emission lines over a broad wavelength range while maintaining high throughput between the lines, which is critical for a wide range of science cases. For example, observations may be made in a very narrow wavelength range between OH emission lines using ultra-narrow band filters \citep{horton2004}. However, this requires a specific unambiguously identifiable feature within the narrow wavelength range from the object. As a result, the number and nature of objects that may be observed by this technique is severely limited. More sophisticated approaches attempt to remove OH emission lines by dispersing the light with a diffraction grating at high resolution, selectively masking out the OH lines, and then recombining the light \citep{iwamuro2001, motohara2002}. Unfortunately, the diffraction grating and the system optics (which the light must pass through twice) unavoidably scatters the bright OH light and the scattered portions cannot be effectively removed. Thus, OH emission lines are better dealt with before the light reaches any dispersing element. There have been attempts to use holographic filters for this purpose. \citet{ouellette2004} demonstrated a device with 10 notches 10\,dB deep and 0.1\,nm wide with 85\% throughput between the notches. Using several of these holographic filters in series would be sufficient to suppress on the order of 100 OH doublets pre-dispersion, but the internotch throughput would be very low in this configuration. See \citet{sce2008} for an in-depth comparison of these OH suppression techniques.

OH suppression using aperiodic fiber Bragg gratings \citep{jbh2004,jbh2008} overcomes many of the shortcomings of these previous approaches and is the best available solution to the NIR sky background problem from the ground \citep{jbh2011a}. Fiber Bragg gratings (FBGs) are common photonic devices widely used in telecommunications. Basic FBGs are single-mode fibers (SMFs) with a periodic refractive index modulation written into the fiber core by exposing it to ultraviolet light. The periodic refractive index modulation induces strong reflections at the Bragg wavelength, $\lambda_{B}=2n_{\mathrm{eff}}\Lambda$, where $\Lambda$ is the spatial period of the refractive index modulation, and $n_{\mathrm{eff}}$ is the effective index of the core. Reflectivities close to 100\% and narrow bandwidths of 0.1\,nm are possible with simple periodic FBGs \citep{ok1999}. 


However, basic periodic FBGs have limited use for OH suppression. Each periodic index modulation can be thought of as producing a single notch. A very large number of periodic index modulations would be required to suppress the dense forest of OH lines in the $J$ and $H$ bands. Writing multiple periodic index modulations in succession in the same fiber results in high loss in the spectral regions between the notches because of  the excessive exposure to ultraviolet light \citep{jbh2004}. Aperiodic FBGs are designed with all notches treated simultaneously resulting in a single complex refractive index modulation that encodes $\approx$\,50 irregularly spaced notches over a span of $\approx$\,100\,nm with high internotch throughput. Figure \ref{fig:afbgresponse} shows the wavelength response of a single aperiodic FBG with 50 notches between 1.47 and 1.58\,$\micron$. 

\begin{figure*}[!ht]
\center
\includegraphics[width=0.75\textwidth,angle=-90]{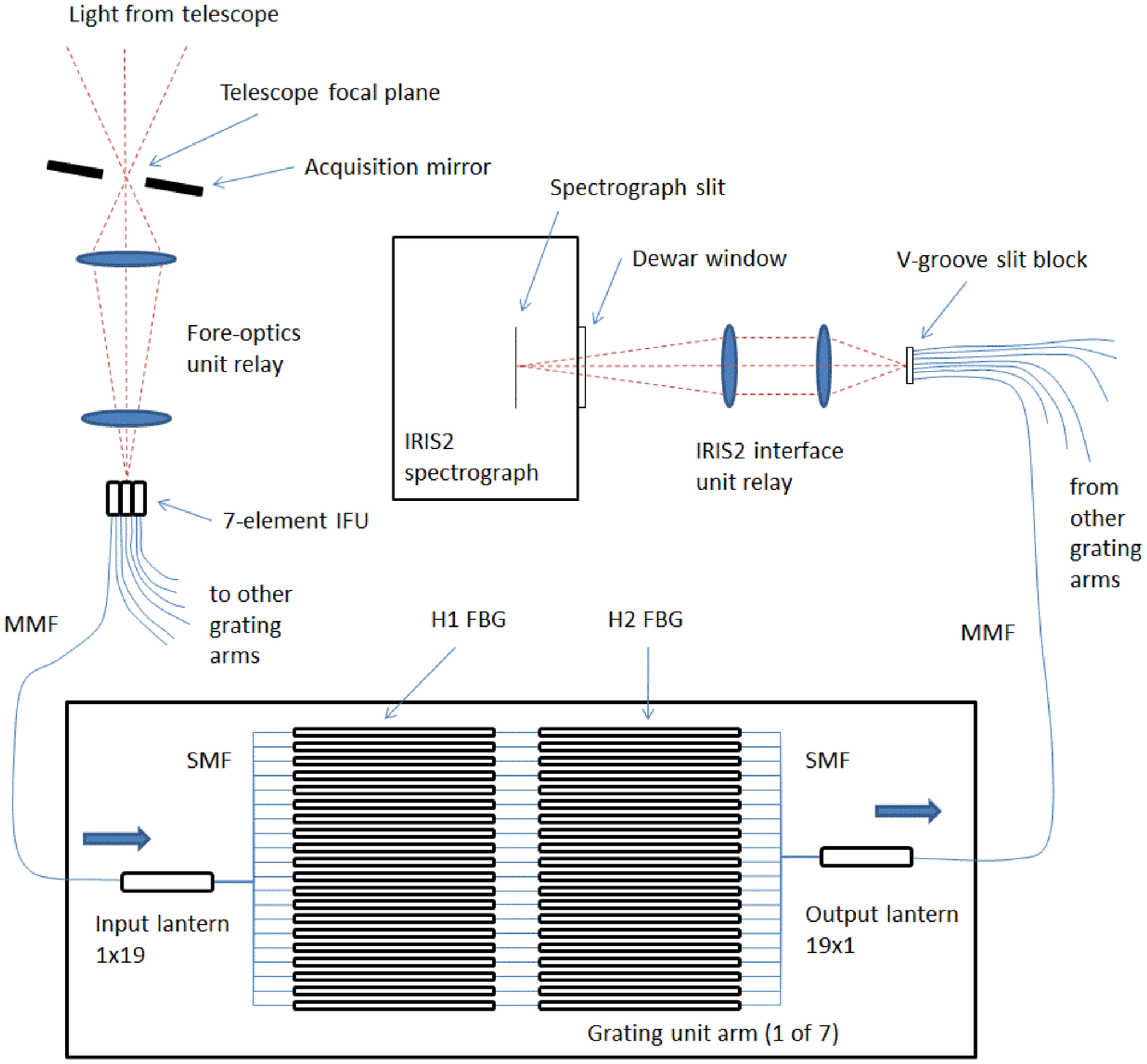} 
\caption{Schematic diagram showing the optical path through each GNOSIS subsystem. Light from the telescope is collected by the fore-optics unit and passed to the grating unit, where OH suppression occurs. The filtered light is passed by another fiber bundle to the IRIS2 interface unit positioned above the IRIS2 dewar window.\label{fig:opticalpath}}
\end{figure*}

Aperiodic FBGs are an attractive means of filtering atmospheric OH lines. However, when used in astronomy, where the wavefront exiting the telescope is distorted by atmospheric turbulence, coupling light into small core diameter SMFs is challenging. As a result, large core diameter multi-mode fibers (MMFs) are more commonly used in astronomy. Unfortunately, MMFs smear out the narrow notches of FBGs into broad, shallow notches because the Bragg condition is different for each fiber mode. OH suppression requires a fiber with the light collecting ability of an MMF and the suppression characteristics of an SMF FBG. 

The solution to this problem is a device called a photonic lantern \citep{sls2005,sls2010,noordegraaf2009,noordegraaf2012}. The device consists of a multi-mode (MM) port connected to an array of SMFs by a taper transition. The photonic lantern converts the modes of the MM port into the supermodes of the SMF array and vice versa. By splicing photonic lanterns to an array of FBGs, we have an ``OH suppression fiber'' that is easy to couple light into and exhibits the exact same transmission characteristics as an FBG in an SMF. 

The GNOSIS grating unit is the first OH suppression unit to utilize these OH suppression fibers. It is independent of telescope and spectrograph but we commissioned the unit at the 3.9\,m Anglo-Australian Telescope (AAT) at Siding Spring Observatory with the existing IRIS2 infrared imaging spectrograph \citep{tinney2004} to demonstrate the potential of OH suppression fibers. In addition to the grating unit, GNOSIS consists of a fore-optics unit and an IRIS2 interface unit, which connects the grating unit to the telescope and spectrograph, respectively. The optical light path is shown in Figure \ref{fig:opticalpath}. Light exiting the back of the AAT is collected by a 7 element integral field unit (IFU) spanning 1.2$\arcsec$ mounted inside the fore-optics unit. The light is transported by a fiber bundle to the grating unit, where OH suppression occurs. The OH suppressed light is transported by another fiber bundle to the IRIS2 spectrograph for measurement. 

In this paper, we present the design and performance of each GNOSIS subsystem and as a whole. We begin by discussing the two components of our OH suppression fibers, the FBGs and the photonic lanterns in Sections \ref{section:fbg} and \ref{section:photoniclantern}. The OH suppression fibers are housed in the GNOSIS grating unit, which is described in Section \ref{section:gratingunit}. In Sections \ref{section:foreoptics} and \ref{section:iris2interface} we discuss the GNOSIS fore-optics unit and IRIS2 interface unit, which connects the grating unit to the telescope and IRIS2, respectively. Section \ref{section:onsky} summarizes the on-sky performance of GNOSIS and Section \ref{section:discussion} contains a discussion on the future of OH suppression fibers.

\begin{figure}[!t]
\center
\includegraphics[width=0.45\textwidth]{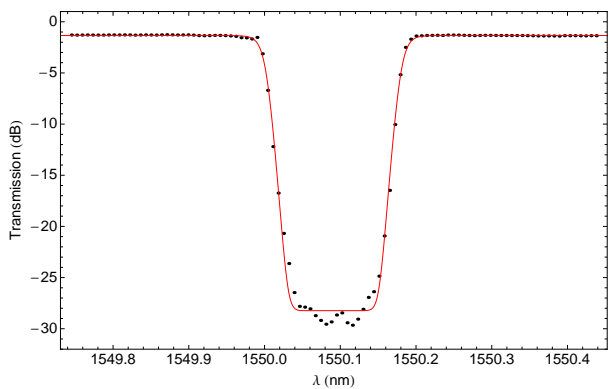}
\caption{Wavelength response (circles) of the H1 FBG showing notch 38 in Table \ref{table:h1notchtable} with the best-fitting Butterworth profile (red line). The notch is deep, narrow, and square, which is ideal for OH suppression.\label{fig:h1notch}}
\end{figure}

\section{Fiber Bragg Gratings}\label{section:fbg}
The GNOSIS OH suppression fibers utilize two complex, multi-notch aperiodic FBGs \citep{jbh2004, jbh2008} in series to suppress OH lines over two-thirds of the $H$ band (1.47--1.7\,$\micron$). The FBGs were manufactured by Redfern Optical Components based on our design to suppress the 103 brightest OH doublets using the line positions and strengths from \citet{rousselot2000}. The H1 FBG has notches in the first portion of the $H$ band (1.47--1.58\,$\micron$) and the H2 FBG has notches in the second portion (1.58--1.7\,$\micron$). The FBGs are printed in a custom photosensitive SMF (NuFern CMS8). FBGs are highly sensitive to both strain and temperature variations, which induce wavelength shifts.  The H1 and H2 FBGs are packaged in 304 stainless steel tubes designed to induce a strain in the fiber that varies with temperature in such a way that it cancels the intrinsic thermal wavelength shift of the FBGs, allowing the notches to remain aligned with the OH lines over the temperature range between -10 and +25\,$^{\circ}$C. 

\begin{figure}[!t]
\center
\includegraphics[width=0.45\textwidth]{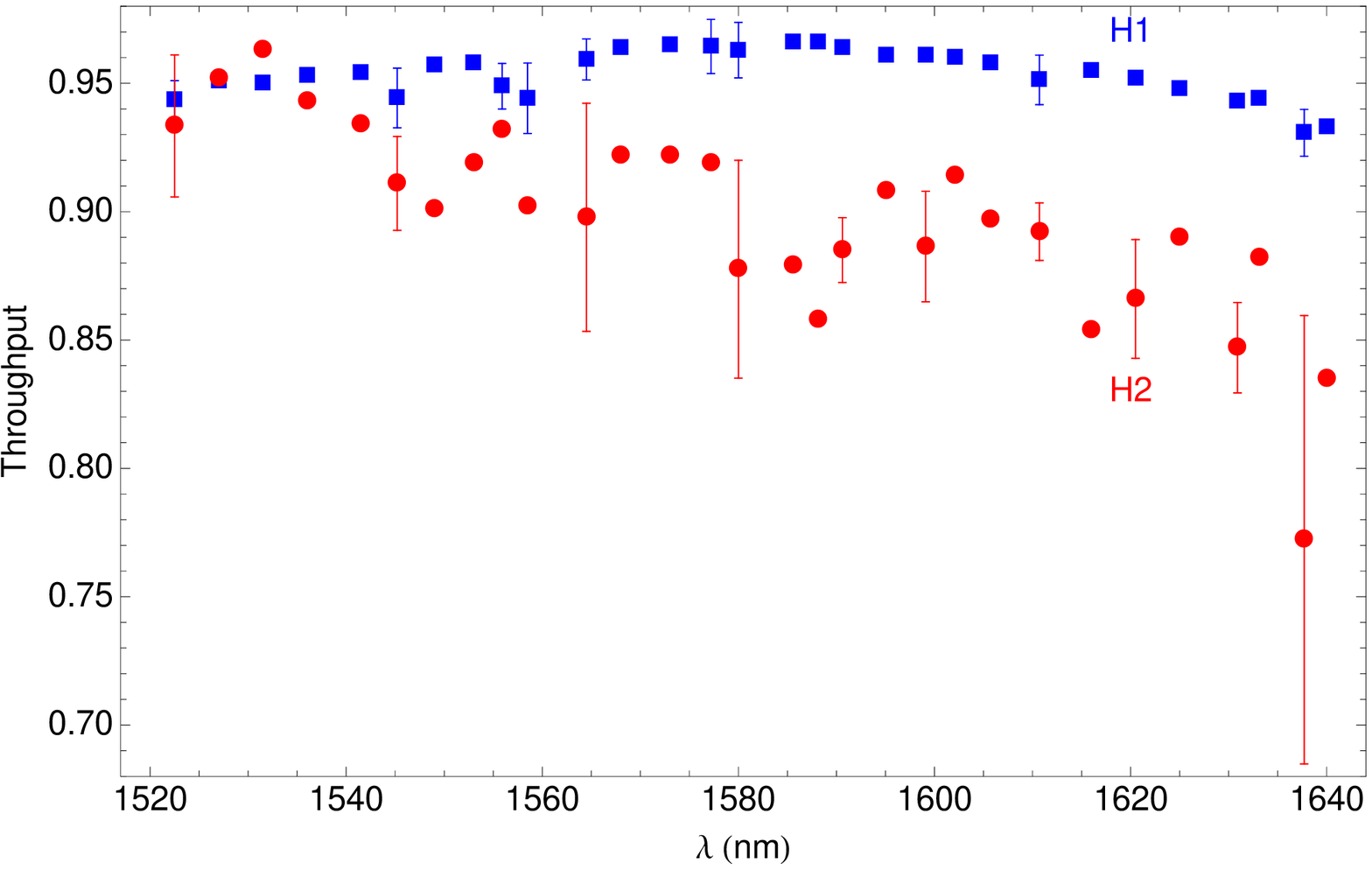}
\caption{Measurements of the H1 (blue squares) and H2 (red circles) FBG internotch throughput using a tunable, narrow external cavity laser. The average throughput (1520--1640\,nm) is 0.954 and 0.896 for the H1 and H2 FBGs, respectively. The throughput of both devices decreases at longer wavelengths mainly to linear loss in the CMS8 fiber. High internotch throughput for the FBGs is critical because these are the regions where scientific observations will be made.\label{fig:internotch_throughput}}
\end{figure}

\begin{table*}[!h]
\caption{H1 FBG Notch Parameters\label{table:h1notchtable}}
\center
\begin{tabular}{cccccccc}
\hline\noalign{\smallskip}
No. & $\lambda_{0}$ & $\sigma(\lambda_{0})$ & $B_{0}-B_{\infty}$ & $\sigma(B_{0}-B_{\infty})$ &$w$ & $\sigma(w)$ & $n$ \\
 & (nm) & (nm) & (dB) & (dB) & (nm) & (nm) &\\
\noalign{\smallskip}\hline\noalign{\smallskip}
 1 & 1466.505 & 0.007 & -30.6 & 4.4 & 0.202 & 0.010 & 8 \\
 2 & 1469.842 & 0.007 & -32.7 & 4.7 & 0.246 & 0.011 & 8 \\
 3 & 1470.220 & 0.007 & -18.2 & 4.1 & 0.186 & 0.008 & 8 \\
 4 & 1471.329 & 0.007 & -13.6 & 3.5 & 0.161 & 0.009 & 8 \\
 5 & 1474.001 & 0.007 & -25.8 & 3.8 & 0.211 & 0.007 & 8 \\
 6 & 1475.555 & 0.005 & -23.5 & 3.2 & 0.167 & 0.012 & 8 \\
 7 & 1477.235 & 0.004 & -28.4 & 3.1 & 0.171 & 0.014 & 8 \\
 8 & 1478.371 & 0.004 & -31.6 & 3.6 & 0.190 & 0.015 & 9 \\
 9 & 1479.982 & 0.005 & -31.7 & 3.9 & 0.320 & 0.018 & 9 \\
 10 & 1480.575 & 0.005 & -28.2 & 3.0 & 0.168 & 0.014 & 7 \\
 11 & 1483.307 & 0.004 & -33.3 & 4.0 & 0.188 & 0.016 & 7 \\
 12 & 1486.438 & 0.004 & -29.7 & 2.7 & 0.143 & 0.013 & 7 \\
 13 & 1488.694 & 0.004 & -28.7 & 3.6 & 0.442 & 0.021 & 7 \\
 14 & 1490.818 & 0.005 & -21.9 & 3.6 & 0.139 & 0.004 & 6 \\
 15 & 1490.984 & 0.005 & -22.4 & 1.4 & 0.146 & 0.015 & 6 \\
 16 & 1493.187 & 0.004 & -30.0 & 2.9 & 0.163 & 0.014 & 9 \\
 17 & 1500.675 & 0.006 & -11.5 & 1.6 & 0.203 & 0.011 & 7 \\
 18 & 1502.516 & 0.005 & -14.4 & 3.0 & 0.118 & 0.006 & 5 \\
 19 & 1502.696 & 0.006 & -14.6 & 1.2 & 0.125 & 0.013 & 5 \\
 20 & 1505.281 & 0.007 & -28.4 & 2.6 & 0.171 & 0.014 & 7 \\
 21 & 1505.553 & 0.007 & -34.4 & 4.6 & 0.171 & 0.013 & 9 \\
 22 & 1506.397 & 0.007 & -22.8 & 2.6 & 0.165 & 0.013 & 7 \\
 23 & 1506.895 & 0.007 & -32.2 & 3.2 & 0.171 & 0.013 & 8 \\
 24 & 1508.222 & 0.008 & -17.7 & 2.4 & 0.163 & 0.013 & 6 \\
 25 & 1508.826 & 0.008 & -28.4 & 2.5 & 0.168 & 0.013 & 8 \\
 26 & 1510.733 & 0.008 & -9.9 & 1.5 & 0.155 & 0.013 & 5 \\
 27 & 1511.369 & 0.008 & -21.9 & 2.4 & 0.163 & 0.012 & 8 \\
 28 & 1514.545 & 0.008 & -13.7 & 1.9 & 0.160 & 0.013 & 6 \\
 29 & 1518.711 & 0.007 & -32.3 & 3.3 & 0.187 & 0.015 & 8 \\
 30 & 1524.092 & 0.007 & -36.2 & 5.6 & 0.202 & 0.016 & 9 \\
 31 & 1528.781 & 0.004 & -34.9 & 4.8 & 0.171 & 0.013 & 9 \\
 32 & 1533.241 & 0.005 & -38.1 & 7.1 & 0.247 & 0.014 & 11 \\
 33 & 1539.536 & 0.004 & -34.2 & 4.5 & 0.173 & 0.013 & 9 \\
 34 & 1543.203 & 0.005 & -38.6 & 8.8 & 0.327 & 0.038 & 8 \\
 35 & 1543.861 & 0.009 & -8.5 & 1.3 & 0.263 & 0.015 & 7 \\
 36 & 1546.214 & 0.005 & -21.8 & 2.2 & 0.177 & 0.012 & 8 \\
 37 & 1547.423 & 0.005 & -16.8 & 1.8 & 0.164 & 0.012 & 8 \\
 38 & 1550.088 & 0.004 & -30.3 & 2.8 & 0.184 & 0.014 & 8 \\
 39 & 1550.979 & 0.004 & -32.0 & 3.5 & 0.182 & 0.014 & 9 \\
 40 & 1551.788 & 0.004 & -25.9 & 2.2 & 0.169 & 0.013 & 8 \\
 41 & 1554.035 & 0.005 & -34.7 & 4.9 & 0.306 & 0.016 & 11 \\
 42 & 1554.615 & 0.004 & -33.8 & 3.9 & 0.201 & 0.014 & 9 \\
 43 & 1557.018 & 0.004 & -30.7 & 2.7 & 0.174 & 0.014 & 8 \\
 44 & 1559.764 & 0.004 & -34.9 & 4.9 & 0.185 & 0.013 & 9 \\
 45 & 1563.134 & 0.004 & -35.0 & 4.5 & 0.193 & 0.015 & 7 \\
 46 & 1565.625 & 0.008 & -32.8 & 3.8 & 0.517 & 0.016 & 12 \\
 47 & 1570.254 & 0.004 & -29.9 & 2.3 & 0.171 & 0.013 & 8 \\
 48 & 1576.030 & 0.006 & -15.6 & 1.2 & 0.220 & 0.011 & 9 \\
 49 & 1578.114 & 0.006 & -17.2 & 2.5 & 0.132 & 0.007 & 7 \\
 50 & 1578.306 & 0.008 & -18.2 & 1.5 & 0.135 & 0.013 & 6 \\
\noalign{\smallskip}\hline
\end{tabular}
\tablecomments{Symbols are defined as follows: $\lambda_{0}$ is the notch center, $B_{0}-B_{\infty}$ is the notch depth, $w$ is the notch width, $n$ is the profile index, and $\sigma$ is the standard deviation in each quantity.}
\end{table*}

\begin{table*}[!h]
\caption{H2 FBG Notch Parameters\label{table:h2notchtable}}
\center
\begin{tabular}{cccccccc}
\hline\noalign{\smallskip}
No. & $\lambda_{0}$ & $\sigma(\lambda_{0})$ & $B_{0}-B_{\infty}$ & $\sigma(B_{0}-B_{\infty})$ &$w$ & $\sigma(w)$ & $n$\\
 & (nm) & (nm) & (dB) & (dB) & (nm) & (nm) &\\
\noalign{\smallskip}\hline\noalign{\smallskip}
 1 & 1583.028 & 0.008 & -28.8 & 1.3 & 0.171 & 0.015 & 7 \\
 2 & 1583.330 & 0.009 & -29.2 & 1.2 & 0.183 & 0.021 & 9 \\
 3 & 1584.252 & 0.007 & -25.5 & 1.9 & 0.157 & 0.014 & 8 \\
 4 & 1584.807 & 0.008 & -29.1 & 1.2 & 0.178 & 0.017 & 8 \\
 5 & 1586.253 & 0.008 & -19.8 & 2.2 & 0.142 & 0.012 & 8 \\
 6 & 1586.932 & 0.009 & -28.2 & 1.2 & 0.180 & 0.016 & 8 \\
 7 & 1589.006 & 0.008 & -12.4 & 1.7 & 0.114 & 0.012 & 8 \\
 8 & 1589.732 & 0.009 & -23.3 & 2.0 & 0.205 & 0.025 & 7 \\
 9 & 1591.590 & 0.011 & -12.2 & 1.6 & 0.473 & 0.018 & 8 \\
 10 & 1592.503 & 0.008 & -2.8 & 0.6 & 0.102 & 0.012 & 5 \\
 11 & 1593.224 & 0.007 & -15.5 & 2.1 & 0.163 & 0.012 & 8 \\
 12 & 1597.263 & 0.009 & -29.0 & 1.1 & 0.195 & 0.017 & 8 \\
 13 & 1597.415 & 0.016 & -4.3 & 1.6 & 0.164 & 0.022 & 8 \\
 14 & 1603.086 & 0.007 & -29.8 & 0.8 & 0.201 & 0.016 & 9 \\
 15 & 1607.975 & 0.007 & -28.5 & 1.3 & 0.161 & 0.015 & 8 \\
 16 & 1612.866 & 0.007 & -30.4 & 0.9 & 0.232 & 0.017 & 10 \\
 17 & 1619.460 & 0.008 & -28.8 & 1.4 & 0.166 & 0.014 & 8 \\
 18 & 1623.539 & 0.007 & -30.9 & 0.9 & 0.270 & 0.017 & 11 \\
 19 & 1627.035 & 0.007 & -11.3 & 1.7 & 0.173 & 0.010 & 8 \\
 20 & 1627.971 & 0.006 & -7.4 & 1.2 & 0.162 & 0.010 & 8 \\
 21 & 1630.229 & 0.007 & -20.9 & 2.5 & 0.196 & 0.013 & 8 \\
 22 & 1631.553 & 0.005 & -15.6 & 4.4 & 0.182 & 0.015 & 9 \\
 23 & 1631.721 & 0.013 & -29.1 & 1.8 & 0.200 & 0.027 & 9 \\
 24 & 1634.178 & 0.008 & -26.5 & 2.2 & 0.211 & 0.014 & 9 \\
 25 & 1635.136 & 0.008 & -30.0 & 1.4 & 0.315 & 0.020 & 10 \\
 26 & 1636.040 & 0.007 & -22.6 & 2.1 & 0.171 & 0.012 & 9 \\
 27 & 1638.854 & 0.007 & -29.1 & 1.6 & 0.208 & 0.013 & 9 \\
 28 & 1641.471 & 0.007 & -25.6 & 2.0 & 0.156 & 0.012 & 9 \\
 29 & 1644.219 & 0.006 & -29.3 & 1.6 & 0.192 & 0.013 & 9 \\
 30 & 1644.765 & 0.007 & -24.3 & 2.5 & 0.222 & 0.013 & 9 \\
 31 & 1647.563 & 0.008 & -25.8 & 2.2 & 0.136 & 0.012 & 7 \\
 32 & 1647.737 & 0.008 & -25.9 & 2.4 & 0.137 & 0.014 & 7 \\
 33 & 1647.907 & 0.007 & -26.6 & 1.9 & 0.144 & 0.015 & 7 \\
 34 & 1650.239 & 0.007 & -29.2 & 1.4 & 0.175 & 0.016 & 8 \\
 35 & 1655.385 & 0.007 & -26.3 & 1.9 & 0.161 & 0.011 & 8 \\
 36 & 1658.635 & 0.007 & -17.5 & 2.2 & 0.237 & 0.011 & 10 \\
 37 & 1661.000 & 0.006 & -19.5 & 2.1 & 0.127 & 0.013 & 7 \\
 38 & 1661.208 & 0.008 & -19.2 & 2.3 & 0.124 & 0.012 & 7 \\
 39 & 1668.914 & 0.007 & -24.7 & 2.3 & 0.137 & 0.014 & 8 \\
 40 & 1669.238 & 0.007 & -30.1 & 1.7 & 0.149 & 0.014 & 7 \\
 41 & 1670.267 & 0.007 & -19.5 & 2.2 & 0.132 & 0.009 & 8 \\
 42 & 1670.882 & 0.007 & -28.2 & 1.7 & 0.149 & 0.015 & 8 \\
 43 & 1672.479 & 0.006 & -15.9 & 1.9 & 0.125 & 0.009 & 7 \\
 44 & 1673.251 & 0.007 & -25.9 & 2.0 & 0.167 & 0.019 & 8 \\
 45 & 1673.391 & 0.006 & -8.3 & 1.9 & 0.192 & 0.011 & 8 \\
 46 & 1675.378 & 0.008 & -10.2 & 1.7 & 0.112 & 0.009 & 8 \\
 47 & 1675.530 & 0.005 & -9.1 & 1.7 & 0.116 & 0.009 & 8 \\
 48 & 1675.623 & 0.007 & -11.7 & 2.0 & 0.119 & 0.002 & 6 \\
 49 & 1676.356 & 0.007 & -20.4 & 2.3 & 0.165 & 0.012 & 7 \\
 50 & 1679.400 & 0.007 & -2.5 & 0.6 & 0.090 & 0.008 & 5 \\
 51 & 1680.237 & 0.007 & -13.7 & 2.1 & 0.161 & 0.011 & 8 \\
 52 & 1684.044 & 0.007 & -27.4 & 1.9 & 0.177 & 0.014 & 8 \\
 53 & 1684.901 & 0.008 & -14.4 & 1.9 & 0.184 & 0.012 & 8 \\
 54 & 1690.362 & 0.008 & -29.6 & 1.7 & 0.191 & 0.017 & 8 \\
 55 & 1695.502 & 0.009 & -27.6 & 2.2 & 0.155 & 0.012 & 8 \\
\noalign{\smallskip}\hline
\end{tabular}
\tablecomments{Symbols are defined as follows: $\lambda_{0}$ is the notch center, $B_{0}-B_{\infty}$ is the notch depth, $w$ is the notch width, $n$ is the profile index, and $\sigma$ is the standard deviation in each quantity.}
\end{table*}

\subsection{Notch Characterization}
We measured the position, depth, and width of each notch from the wavelength response of the H1 and H2 FBGs ($R\sim$\,10,000). The FBGs were not designed with any particular functional form for the notch profile in mind, but we found that a Butterworth profile provided a good fit to the empirical data. Thus, each notch in the wavelength response was fitted with a Butterworth profile,
\begin{equation}
B(\lambda)=B_{\infty}-\frac{B_{\infty}-B_{0}}{1+\left[\frac{2(\lambda_{0}-\lambda)}{w}\right]^{2n}},
\end{equation}
where $\lambda_{0}$ is the profile center, $B_{0}$ is the profile value at the center, $w$ is the profile width, $B_{\infty}$ is the profile value far away from the profile center, and $n$ is the profile index. The value of $n$ affects the steepness of the profile. From the Butterworth parameters for each notch, we take $\lambda_{0}$ to be the notch position, $B_{0}-B_{\infty}$ to be the notch depth, and $w$ to be the notch width. Figure \ref{fig:h1notch} shows one of the notches in the H1 FBG and the best-fitting Butterworth profile (red line). 

The H1 FBG notch parameters are listed in Table \ref{table:h1notchtable}. The H1 FBG notch parameters are averages measured from the wavelength response of eight separate devices obtained by scanning a narrow tunable laser source across the device bandwidth and comparing with a reference fiber. A total of 50 notches were measured in the range 1465--1580\,nm in each of the eight H1 devices. The H2 FBG notch parameters are listed in Table \ref{table:h2notchtable}. The H2 FBG notch parameters are averages measured from the wavelength response of 106 separate devices obtained using a broadband source and an optical spectrum analyzer. A total of 55 notches were measured in the range 1580--1700\,nm in each of the 106 H2 devices. Tables \ref{table:h1notchtable} and \ref{table:h2notchtable} contain more notches than the original design of 103. The additional notches may be printing errors from the manufacturing process. For example, notch 9 in Table \ref{table:h2notchtable} is not associated with any OH line from \citet{rousselot2000}, but we include it here because it is a significant spectral feature of the FBGs. Also, some notches are deliberately designed to be wider in order to suppress closely spaced OH lines with a single notch.

\begin{figure}[!t]
\center
\includegraphics[width=0.45\textwidth]{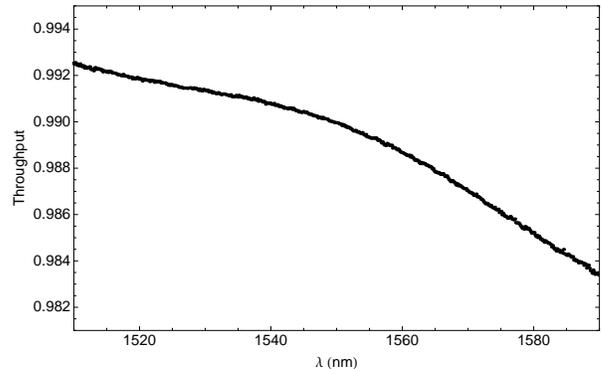}
\caption{Throughput of 1\,m of NuFern CMS8 fiber measured using a cutback technique. The decrease in throughput at longer wavelengths results in lower internotch throughput for the H1 and H2 FBGs at longer wavelengths.\label{fig:cms8_loss}}
\end{figure}

\subsection{Throughput}
The throughput of the FBGs between notches is of interest because this is the region where scientific observations are made. The throughput at several internotch wavelengths was measured using a cutback technique with a Photonetics Tunics tunable external cavity laser (1520--1640\,nm) and a Thorlabs PM100D power meter with an S122C sensor head. The laser source was calibrated to $\pm$0.2\,nm before the measurements were carried out. The internotch measurement wavelengths were chosen to avoid any notches. Based on the values listed in Tables \ref{table:h1notchtable} and \ref{table:h2notchtable} the closest any internotch measurement value comes to the center of a notch is $0.65\pm0.2$\,nm. Given that the average notch width is $\approx\,0.2$\,nm, the internotch measurements should have safely avoided all FBG notches. 

The internotch throughput of the H1 and H2 FBGs in the range 1520--1640\,nm are shown in Figure \ref{fig:internotch_throughput}. The average throughput is 0.954 and 0.896 for the H1 and H2 FBGs, respectively. A total of nine H1 FBGs were randomly selected and measured. The points with error bars show the average internotch throughput of all nine H1 FBGs and the 1$\sigma$ variation. The points without error bars are the measured values from one of the nine H1 FBGs. A total of five H2 FBGs were randomly selected from an initial delivery of 19 devices and measured. Again, the points with error bars show the average and 1$\sigma$ variation of all five H2 FBGs and points without error bars are measurements from one of the 5 devices. The throughput in both devices decreases at longer wavelengths, but this is due mainly to the decrease in the transmission of the CMS8 fiber at longer wavelengths (see Figure \ref{fig:cms8_loss}).

The performance of the H1 and H2 FBGs are highly satisfactory for OH suppression. On average the notches are deep (24\,dB), narrow (0.19\,nm), and square and the average throughput between the notches is high (0.92). 

\section{Photonic Lanterns}\label{section:photoniclantern}
The GNOSIS OH suppression fibers use two 1$\times$19 photonic lanterns (one for input and one for output) to provide the fibers with the light collecting ability of an MMF. The photonic lanterns were manufactured by NKT Photonics \citep{noordegraaf2012} and consist of a 50\,$\micron$ core diameter MM port connected to an array of 19 SMFs by a taper transition with a taper ratio of 11 enclosed in a protective metal case. A 5\,m long FC/PC-connectorized delivery fiber is fusion spliced to the MM port to facilitate connecting the photonic lanterns with other components. The SMFs are approximately 3\,m in length and not connectorized. 

The photonic lanterns are designed to efficiently convert the modes of the MM port into the supermodes of the SMF array and vice versa.  The SMF array will only support $N$ supermodes independent of wavelength, where $N$ is equal to the number of SMFs. Thus, for maximum efficiency the MM port should be designed to support $M=N$ modes. However, the number of modes supported by the MM port is wavelength dependent ($M(\lambda)\propto \lambda^{-2}$). For GNOSIS the MM port is designed to support $M=N=19$ modes at $\lambda=1.55\,\micron$ with a $d=50\,\micron$ core diameter, but there will be more (fewer) modes at shorter (longer) wavelengths.



\subsection{Numerical Aperture}\label{section:numericalaperture}
The MM port parameters determine the numerical aperture (NA) by \citep{noordegraaf2012}
\begin{equation}\label{equation:lanternNA}
\mathrm{NA}\approx\frac{2\lambda\sqrt{N}}{\pi d}.
\end{equation}
Equation (\ref{equation:lanternNA}) gives an NA of 0.086 ($f$/5.8) for the GNOSIS photonic lanterns. We cannot measure this value directly because of the delivery fiber fusion spliced to the MM port. Instead, we measured the focal ratio degradation (FRD) in the delivery fiber and computed the MM port NA from the measured output NA of the delivery fiber. We fusion spliced an amplified spontaneous emission (ASE) source centered at 1.53\,$\micron$ to a randomly chosen SMF. The NA of the light exiting the delivery fiber was taken to be the value that enclosed 95\% of the total energy in the far-field image. The measurement was repeated 10 times for each photonic lantern. The NA of the light exiting the delivery fiber is 0.103$\pm$0.005 on average for the 14 photonic lanterns. The laboratory measurement of the delivery fiber FRD ratio gave $f_{\mathrm{in}}$/$f_{\mathrm{out}}\approx 1.2$. Thus, the average value of the MM port NA is 0.085$\pm$0.004, which is in agreement with the estimate from Equation \ref{equation:lanternNA}. The measured MM port NA values for each photonic lantern is listed in Table \ref{table:plmeasurements}.

For minimum loss in the MM to SM conversion, the MM port must be fed at $>$ $f$/5.8 in order to avoid overfilling. To account for FRD in the delivery fiber the OH suppression fibers should be fed at $>$ $f$/7 to ensure the light arrives at the MM port at $>$ $f$/5.8. At the output end of the OH suppression fibers, it is important to know the f-ratio of the beam exiting our fibers to properly interface them with the subsequent components. Based on our measurements of the output photonic lanterns the beam exiting our OH suppression fibers is $\approx$\,$f$/4.9.

\begin{figure*}[!t]
\center
\includegraphics[width=0.5\textwidth,angle=-90]{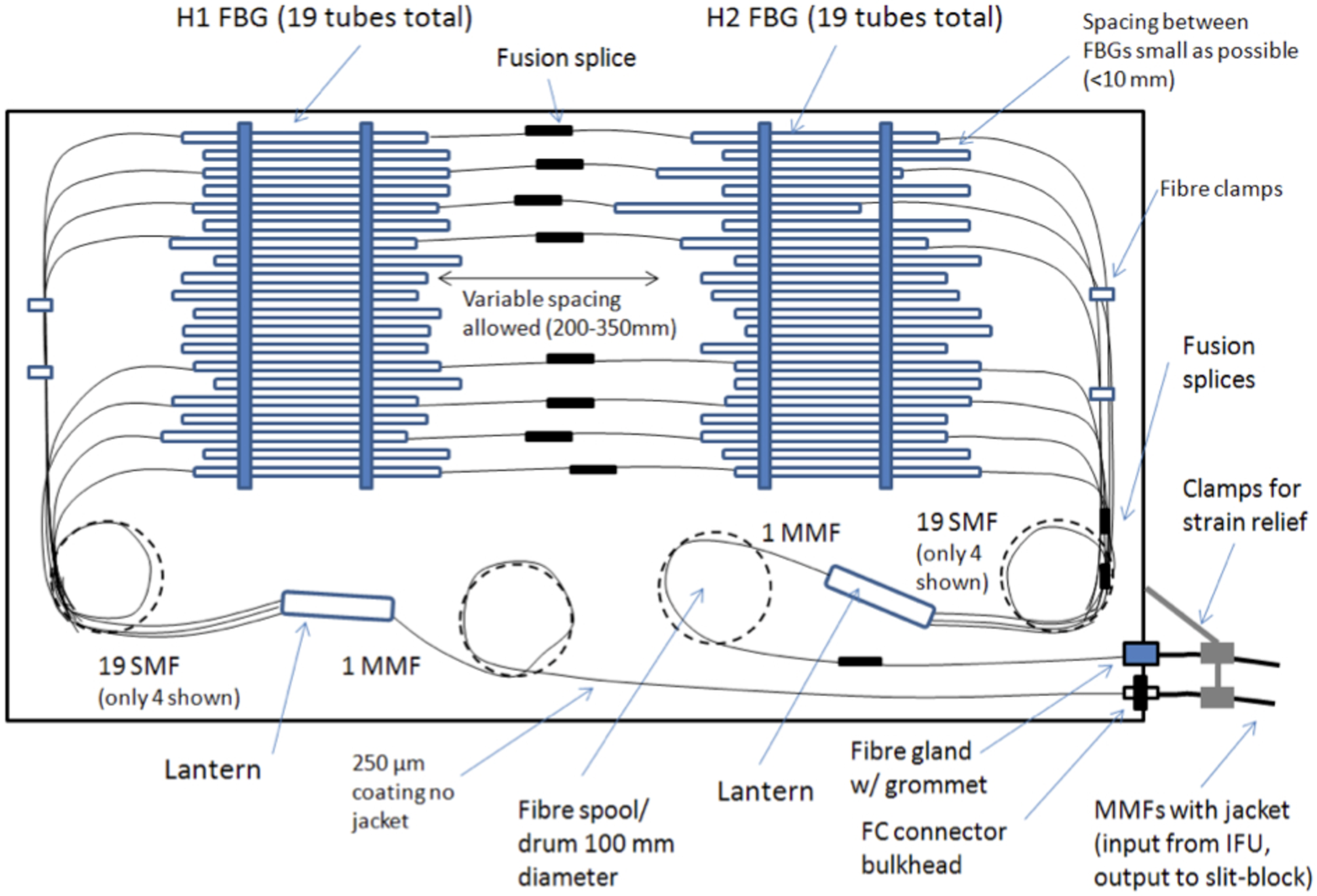}
\caption{Schematic diagram of one of the OH suppression fibers of the GNOSIS grating unit. Light enters the MMF of the input photonic lantern and transitions to 19 SMFs. Light in the 19 SMFs is filtered by an H1 and an H2 FBG. The 19 SMFs are recombined into a single MMF using an output photonic lantern.\label{fig:gratingunit01}}
\end{figure*}

\subsection{Throughput}
At the design wavelength, the number of modes supported by the MM port exactly matches the number of supermodes of the SMF array resulting in minimum loss in the taper transition. We determined the SM to MM throughput by splicing the 1.53\,$\micron$ ASE source to a randomly chosen SMF and measured the output power of the delivery fiber. The measurement was repeated 10 times for each photonic lantern. The average SM to MM throughput of the seven output photonic lanterns is 0.974$\pm$0.009 with the best throughput at 0.989. 


We measured the MM to SM throughput by injecting a beam with an NA of 0.086 ($f$/5.8) from an ASE source centered at 1.53\,$\micron$ into the delivery fiber and collecting the power from all 19 SMFs. The average MM to SM throughput of the five measured input photonic lanterns is 0.851$\pm$0.0095. Recall previously that we found that the delivery fiber must be fed at $>$ $f$/7 to compensate for FRD. Hence, in our measurement scheme we were overfilling the MM port, causing the throughput to be significantly lower than the SM to MM throughput. However, the GNOSIS fore-optics unit feeds the OH suppression fibers at $\approx$\,$f$/5 and the measured MM to SM throughput gives a sense of the performance under operating conditions. For reference, the throughput of the photonic lanterns near the design wavelength is listed in Table \ref{table:plmeasurements}.

\begin{table}[!t]
\caption{Photonic Lantern Properties for each OH Suppression Fiber at 1.53\,$\micron$\label{table:plmeasurements}}
\center
\begin{tabular}{cccc}
Input & Input & Output & Output\\
\noalign{\smallskip}\hline\noalign{\smallskip}
MM to SM & NA  & SM to MM & NA  \\
Throughput & & Throughput & \\
\noalign{\smallskip}\hline\noalign{\smallskip}
0.857 & 0.081 &  0.996 & 0.084  \\
0.849 & 0.082 &  0.975 & 0.089  \\
0.838 & 0.078 &  0.962 & 0.088  \\
Not measured & 0.089&  0.973 & 0.088  \\
0.863 & 0.088 &  0.984 & 0.084  \\
0.849 & 0.088 &  0.973 & 0.088  \\
Not measured & 0.091 &  0.989 & 0.079  \\
\noalign{\smallskip}\hline{\smallskip}
\end{tabular}
\end{table}

We have only measured the throughput of the photonic lanterns near the design wavelength. In theory, the throughput of the photonic lantern depends on wavelength because the number of modes supported by the MM port depends on wavelength. In the suppression range of GNOSIS (1.47--1.7\,$\micron$), the number of modes varies from 21 to 16, ignoring dispersion. In the MM to SM conversion, the throughput is lower at the blue end because there are more modes in the MM port than SMFs. In the SM to MM conversion, the throughput is lower at the red end because there are more SMFs or supermodes than can be supported by the MM port. Thus, when two identical photonic lanterns are used back-to-back, the loss at the blue end during the MM to SM conversion and the loss at the red end in the SM to MM conversion should produce a wavelength response that is peaked at the design wavelength. Although this behavior has not been demonstrated for a symmetric photonic lantern system alone, the effect is present in an integrated OH suppression fiber. The wavelength response of an OH suppression fiber is discussed in Section 4.3 and the throughput appears to be peaked at $\approx$\,1.53\,$\micron$. 

\section{Grating Unit}\label{section:gratingunit}
The GNOSIS grating unit contains seven independent OH suppression fibers. Each OH suppression fiber consists of a 1$\times$19 input photonic lantern, 19 pairs of H1+H2 FBGs, and an output 19$\times$1 photonic lantern all fusion spliced together and arranged on a plastic tray as shown in Figure \ref{fig:gratingunit01}. The seven trays are mounted within an aluminum enclosure and padded with foam sheets. The grating unit enclosure is mounted in a standard equipment rack within the Cassegrain cage during use.

\subsection{Splice Losses}
In assembling the OH suppression fibers, all splices were carried out using a Fitel S175 v.2000 fusion splicer. CMS8 to CMS8 fusion splices were carried out using the standard identical SMF prescription, which results in an average splice loss of $\approx$\,0.05\,dB. CMS8 to SMF-28 fusion splices were carried out using a custom prescription, which results in an average splice loss of $\approx$\,0.15\,dB. All splices are protected by splice sleeves and excess fiber is spooled into circular drums in each tray. CMS8 fiber is much lossier than SMF-28 so we minimized the length of CMS8 fiber as much as possible during the assembly of the OH suppression fibers. 

\begin{figure*}[!ht]
\center
\includegraphics[width=1\textwidth]{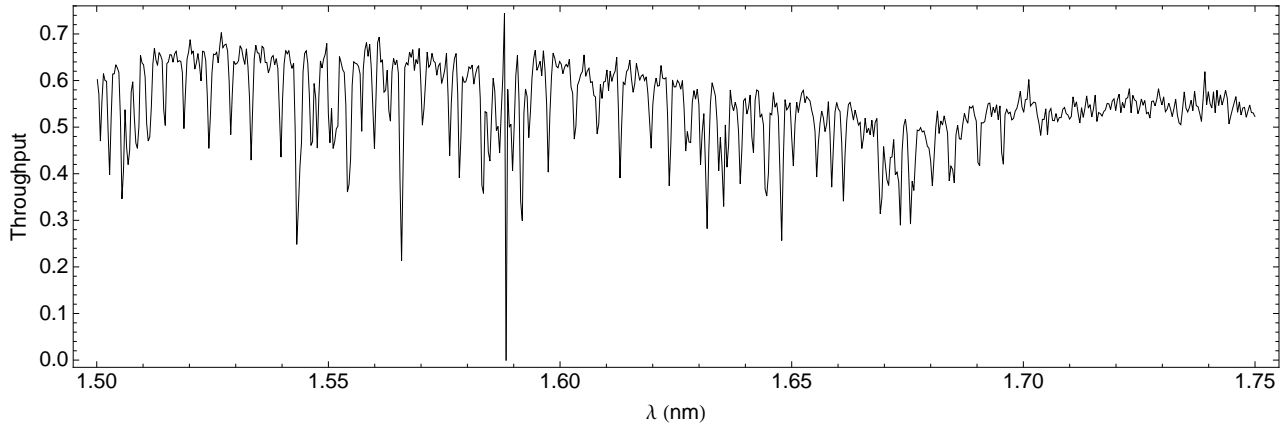}
\caption{Throughput of GNOSIS OH suppression fiber 5 (central sky fiber) measured on-telescope during 2011 September commissioning run by comparing spectra with and without OH suppression. The internotch throughput is relatively high and the FBG notches are evident although not as deep as reported in Tables \ref{table:h1notchtable} and \ref{table:h2notchtable} due to the low resolution of the IRIS2 spectrograph.\label{fig:gratingthroughputvswavelength}}
\end{figure*}

\subsection{Throughput}

We measured the throughputs of all seven OH suppression fibers before installation during the first commissioning run by simulating illumination by the telescope with a Thorlabs S5FC1550S(P)-A2 SLD source centered at 1.55\,$\micron$. The fibers were fed with a $\approx$\,$f$/5 beam by butt-coupling a MMF to the fiber using a FC-to-FC connector and the output power was measured using a Thorlabs PMD100D power meter with an S122C sensor head. The measured values are listed in Table \ref{table:labthroughputmay2011}. The average throughput of the seven OH suppression fibers is $0.58\pm0.05$ with a maximum and minimum of 0.61 and 0.55, respectively.

We may compare this measurement to an estimate based on the measurements of individual components. From the CMS8 to SMF-28 splice loss (0.15\,dB), CMS8 to CMS8 splice loss (0.05\,dB), the average MM to SM throughput (0.851) of the input photonic lanterns at f/5.8, the average SM to MM throughput (0.974) of the output photonic lanterns, and the average internotch throughput of the H1 (0.954) and H2 (0.896) FBGs in the range 1520--1640\,nm, the estimated throughput of an OH suppression fiber is $\approx$\,0.61 including Fresnel reflections from the FC-to-FC connector. Additionally, we must adjust the MM to SM throughput to an $f$/5 input beam. Based on laboratory measurements of a single photonic lantern, the MM to SM throughput for an $f$/5 beam is $\approx$\,0.25\,dB lower than an $f$/5.8 beam, i.e., 0.803. With this adjustment, the estimated throughput is 0.575, which is in agreement with the measured value for the integrated OH suppression fiber. 

\subsection{Wavelength Response}
Figure \ref{fig:gratingthroughputvswavelength} shows the wavelength response of OH suppression fiber 5. The wavelength responses of the OH suppression fibers were measured after installation and alignment of the fore-optics unit (see Section \ref{section:foreoptics}) and IRIS2 interface unit (see Section \ref{section:iris2interface}). IRIS2 was used to obtain a spectrum of the dome flat lamp through the entire system with and without an OH suppression fiber. The ratio of the two spectra gives the wavelength response of the OH suppression fiber. The notches of the FBGs appear as significant dips in the throughput, but they are not as deep or narrow as indicated in Tables \ref{table:h1notchtable} and \ref{table:h2notchtable} because the resolution of IRIS2 is not high enough to resolve these notches. The internotch throughput is relatively high near the design wavelength of the photonic lanterns and there is some dependence on wavelength due to the linear loss of the CMS8 and the wavelength-dependent loss from a symmetric photonic lantern system.

\section{Fore-optics Unit}\label{section:foreoptics}
The grating unit is interfaced with the AAT using a fore-optics unit that is mounted at the AAT Cassegrain focus. The main function of the fore-optics unit is to re-image the central region of the AAT focal plane onto an IFU and feed this light to the grating unit. The first optical element encountered by the $f$/8 beam from the AAT is an acquisition mirror. The acquisition mirror is just a circular mirror with a central aperture. The central portion of the beam passes through the aperture to the IFU while the rest of the beam is diverted to an acquisition camera. 

The portion of the beam that passes through the acquisition mirror is then magnified to $f$/265 by an optical relay consisting of a magnifying achromatic doublet lens and a doublet field-flattening lens with a pupil stop positioned in the telescope pupil plane between the two lenses. Each lens has an AR coating with a reflectance $<1$\% over the waveband from 1.0 to 1.7\,$\micron$. 

\begin{figure}[!t]
\center
\includegraphics[width=0.4\textwidth]{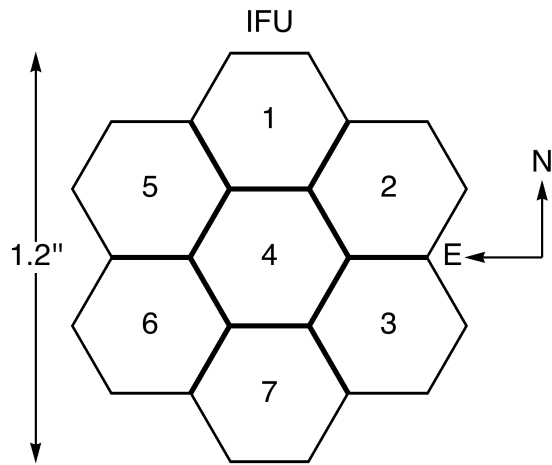} 
\caption{Diagram of the GNOSIS IFU orientation on the sky. The fore-optics relay gives a plate scale of 0.2$\arcsec$\,mm$^{-1}$ at the IFU surface. Each 2\,mm flat-to-flat hexagonal lenslet spans 0.4$\arcsec$ and the total IFU spans 1.2$\arcsec$. The numbering indicates the IRIS2 interface unit slit element connected to each IFU element.\label{fig:ifuslit}}
\end{figure}

The beam is then captured by an IFU consisting of an array of seven hexagonal lenslets made of fused silica arranged as shown in Figure \ref{fig:ifuslit}. The front face of each lenslet is polished and AR-coated. The lenslets are glued onto a fused silica substrate, which is fixed in a mount attached to the optical relay assembly. At the IFU surface the plate scale is 0.2$\arcsec$\,mm$^{-1}$ due to the magnification of the optical relay. Thus, each 2\,mm flat-to-flat lenslet spans 0.4$\arcsec$ on the sky and has a field of view (FOV) of 0.14\,arcsec$^{2}$. The total IFU spans 1.2$\arcsec$ on-sky, which is about the median seeing in the $H$ band at Siding Spring Observatory, and the total IFU FOV is 0.97 arcsec$^{2}$.

Each lenslet feeds an $f$/6.5 beam into a 50\,$\micron$ core diameter MMF positioned at the lenslet focus at the back of the IFU substrate. The MMFs are approximately 5\,m in length and feed a $\approx$\,$f$/5 beam to the OH suppression fibers in the grating unit, based on an FRD estimate from \citet{poppett2010} and laboratory measurements. The fibers are jacketed, reinforced, and bundled together in a limited-bend armoured conduit. At the IFU end, the fibers are terminated with a glass ferrule and were attached to the substrate using a UV curing adhesive (Norland Optical Adhesive 61) prior to potting with a polyeurethane encapsulant (Opti-tec 4200). During the attachment process each fiber was individually aligned with the corresponding lenslet by maximizing the power coupled into the fiber from a test source simulating the illumination of the fore-optics by the telescope. At the grating unit end, the fibers are terminated in FC/PC connectors 


\subsection{Throughput}
The throughput of the fore-optics unit was measured during assembly and installation by illuminating the entrance aperture with an $f$/8 beam from a Thorlabs S5FC1550S(P)-A2 SLD source to simulate illumination by the AAT. Before the IFU was installed, the power at the IFU position was measured using a Thorlabs PM100D power meter with an S122C sensor head and compared to the power at the entrance aperture yielding a throughput of $0.865\pm0.05$ for the fore-optics relay. 

After the IFU was installed, a Xenics Xeva-1.7-640 camera was used to image the front and back surfaces of the IFU. The fraction of incident power captured by each IFU element was estimated from these images, correcting for aperture losses at the array and non-uniformity in the illumination from the SLD source. Comparing the input power at each IFU element, which is the fraction of incident power captured by each IFU element times the output power of the relay, to the power at the end of each fiber gives the IFU (plus fiber bundle) throughput of each element. The average IFU throughput is $0.83\pm0.12$. Thus, the average total fore-optics unit throughput (relay+IFU+fiber bundle) is $\approx$\,0.72. The individual throughput of each fore-optics unit element is listed in Table \ref{table:labthroughputmay2011}. 




\section{IRIS2 Interface Unit}\label{section:iris2interface}
The $\approx$\,$f$/4.9 beams from the GNOSIS OH suppression fibers are fed to the IRIS2 infrared imaging spectrograph by 50\,$\micron$ core diameter MMFs 12\,m in length. For these observations, IRIS2 was positioned below the horseshoe of the telescope mount on the dome floor. The fibers are jacketed and enclosed together in a steel coil limited bend armored conduit. At the grating unit end, the fibers are FC/PC-connectorized and are butt-coupled to the OH suppression fibers using FC-to-FC connectors. At the IRIS2 end, the fibers are terminated in a linear slit block consisting of a 7-element fused silica V-groove array from Ocean Optics.  The grooves are designed to hold the fibers, which have a 125\,$\micron$ diameter cladding, with a center-to-center spacing of 250\,$\micron$. The fibers are glued to the grooves and they are encased in epoxy with a protective casing at the back end for strain relief. The slit block is mounted to the IRIS2 interface unit assembly, which sits over the IRIS2 dewar window. The interface unit assembly rests on the structure usually used to attach IRIS2 to the telescope via kinematic mounts and includes various adjustment mechanisms to align the slit block with the slit masks in the IRIS2 slit wheel. 

The interface unit includes an optical relay consisting of two achromatic doublet lenses with a magnification of 3 that images the fibers onto the IRIS2 slit plane, which is inside the instrument dewar. Estimating the FRD in the fiber bundle using \citet{poppett2010}, we expect an $\approx$\,$f$/4 beam from the fiber in the slit block which is magnified to $\approx$\,$f$/12 and the fibers are 150\,$\micron$ in diameter at the slit plane. The magnification yields a resolving power of $R\approx$\,2350 for IRIS2 and satisfies the maximum acceptance cone of IRIS2 ($f$/8).

No modifications were made to IRIS2 except the addition of custom cold stops and slit masks. Two cold stops ($f$/10 and $f$/12) are available for use with GNOSIS. Three slit masks were made by laser drilling with a slit of seven linear 250/200/180\,$\micron$ diameter holes spaced 750\,$\micron$ apart in the same position in the spectral axis as for the standard $H$-offset slit wheel. 

\subsection{Throughput}
The throughput of the IRIS2 interface unit was measured before installation by simulating illumination by the AAT. The entire GNOSIS system was connected and the entrance aperture of the fore-optics unit was illuminated with an $f$/8 beam from a Thorlabs S5FC1550S(P)-A2 SLD source. We measured the output power of each OH suppression fiber in the grating unit and the power in the slit image when the IRIS2 interface unit is connected to the grating unit. The average throughput of the IRIS2 interface unit is 0.93$\pm$0.05 and the measured value for each element is listed in Table \ref{table:labthroughputmay2011}. 



\subsection{Slit Block Alignment}
There are additional losses from the alignment of the slit block with the slit mask. Even with perfect alignment, low-level aberrations will result in a loss of a few percent depending on the size of the slit mask holes and the focal ratio out of the fibers in the slit block. First, focus was adjusted with the dome flat lamp until the central fiber spot image had an FWHM of $\sim1.8$\,pixels. Next, tip-tilt was adjusted followed by an alignment of the central fiber to within 0.1\,pixels of the central slit mask hole. Then, in-plane rotation was adjusted until the position of the outer fiber images are within 0.1\,pixels of each other. The central fiber was again aligned to the central slit mask hole. Lastly, small translational adjustments were then made to maximize the throughput. The throughput numbers vary from alignment to alignment, but it was typically above 90\% and around 97\% for the central fiber with the 250\,$\micron$ slit. There was a noticeable drop by a few percent in the throughput away from the central fiber, indicating that the magnification of the relay is slightly too large. This was confirmed by comparing the difference between the spot positions and the slit mask hole positions for all seven fibers. In calculations we assume an average throughput of 0.95 for slit alignment. 



\begin{table}
\center
\caption{Throughput of each GNOSIS Fiber at 1.55\,$\micron$\label{table:labthroughputmay2011}}
\begin{tabular}{ccccc}
\hline\noalign{\smallskip}
Sky Pos & Fore-optics  & Grating Unit & IRIS2 Interface & Overall \\
\noalign{\smallskip}\hline\noalign{\smallskip}
 N & 0.790 & 0.612 & 0.91 & 0.38 \\
 WNW & 0.727 & 0.579 & 0.93 & 0.35 \\
 WSW & 0.849 & 0.563 & 0.93 & 0.37 \\
 C & 0.752 & 0.596 & 0.93 & 0.42 \\
 ENE & 0.752 & 0.583 & 0.93 & 0.43 \\
 ESE & 0.548 & 0.551 & 0.92 & 0.42 \\
 S & 0.605 & 0.551 & 0.93 & 0.31 \\
 \noalign{\smallskip}\hline\noalign{\smallskip}
 Average & 0.718 & 0.576 & 0.93 & 0.38\\
\noalign{\smallskip}\hline{\smallskip}
\end{tabular}
\end{table}


\section{On-Sky Performance}\label{section:onsky}

We summarize the on-sky performance of GNOSIS below. An in-depth analysis of the on-sky performance of GNOSIS may be found in \citet{sce2012}. GNOSIS was commissioned at the AAT over five separate observing runs spread over the months of March, May, July, September, and November of 2011. Initially, the fore-optics unit was connected to the grating unit by butt-coupling with FC-to-FC connectors. From the May commissioning run onward, the fore-optics unit was fusion spliced directly to the OH suppression fibers within the grating unit. The system was configured to maximize the throughput of central sky fiber. With the remaining fibers we maximized the average throughput of the system by connecting the highest throughput fibers with each other. However, this results in greater fiber-to-fiber throughput variation. In most observations one fiber was configured to bypass the grating unit to serve as a control fiber without OH suppression. 

\subsection{Data Reduction}
The data reduction procedure is similar to the procedure described in \citet{sce2012}. Spectroscopic observations were made with the $H$ broadband filter, $f$/12 coldstop, and the slit mask with 180 or 250\,$\micron$ diameter holes in multiple-read mode (MRM). In MRM, the detector read noise is minimized ($\approx$\,8\,$e^{-}$) because the 1024$\times$1024 Rockwell Hawaii-1 detector is non-destructively read out during the exposure and the final image is a linear least-squares fit through all the reads. The detector dark current is $\approx$\,0.0015\,$e^{-}$ s$^{-1}$. 

Each image was corrected for detector non-linearities and the spectrum in each fiber was extracted using ``Gaussian summation extraction by least squares'' from \citet{sb2010}. 

Each spectrum was corrected for fiber-to-fiber throughput variations measured from observations of the dome flat lamp. The dome flat lamp spectrum in each fiber was extracted and integrated. The values were normalized to the mean value of all seven fibers to give the fiber-to-fiber throughput variation. The measured fiber-to-fiber variation is consistent with the laboratory values listed in Table \ref{table:labthroughputmay2011}. 

The wavelength calibration for each fiber was determined from a xenon arc lamp observation. The arc lamp spectrum in each fiber was extracted and the pixel position of each xenon line was fit by a cubic polynomial. 

The spectra were also corrected for inter-quadrant cross-talk \citep{tinney2003} because the spectra span two quadrants of the IRIS2 detector. In each spectrum, the counts at wavelengths below the $H$ filter cutoff (1.5\,$\micron$) come from the detector. After correcting for detector dark current, the median count rate at $\lambda<1.47\,\micron$ was subtracted from the spectrum.

The final correction we applied to all observations was an instrument response correction, which corrects for the variation in throughput with wavelength. The instrument response was measured from A0V standard star observations. The spectrum in each fiber was extracted and reduced as discussed above. Then the instrument response for each fiber was taken to be the sky-subtracted spectrum divided by a model spectrum of Vega \citep{ck1994} normalized to the median value in the range 1.5--1.69\,$\micron$. 


\subsection{Throughput}
The average throughput of GNOSIS (fore-optics + grating unit + IRIS2 relay) was measured to be $\approx$\,0.38 in a laboratory setting at 1.55\,$\micron$ (see Table \ref{table:labthroughputmay2011}). On-sky, we must include the additional losses from the telescope, the slit block alignment, the IRIS2 spectrograph, and aperture losses for a point source. The throughput of the AAT is $\approx$\,0.88 including reflections from the two mirrors and extra loss for the accumulation of dirt on the mirrors. The throughput of IRIS2 is $\approx$\,0.12 \citep{sce2012}. For the median seeing of 1.2$\arcsec$, the aperture loss is approximately 0.3--0.5 for a point-source offset between 0--0.6$\arcsec$ from the center of the hexagonal array \citep{sce2012}. Thus, the end-to-end throughput of our system was expected to be $\approx$\,0.04 for a diffuse source and $\approx$\,0.018 for a point source with typical aperture losses. 

We measured the end-to-end throughput of the system from observations of A0V stars taken during the September commissioning run, which had the best observing conditions. The spectra were combined by summing each spectral pixel, neglecting the difference in the wavelength calibration of each fiber (less than 1.5\,pixels over most of the detector). The combined spectrum of the A0V star was divided by a model Vega spectrum scaled to the appropriate brightness and assuming a value for the aperture losses. The $I$-band seeing was measured to be $\approx$\,1.5$\arcsec$. Although we are only able to estimate the aperture losses for this seeing, the A0V star observations were consistent with a throughput of $\approx$\,0.02--0.04 in agreement with the expected value from laboratory measurements. 

\begin{figure}
\center
\includegraphics[width=0.45\textwidth]{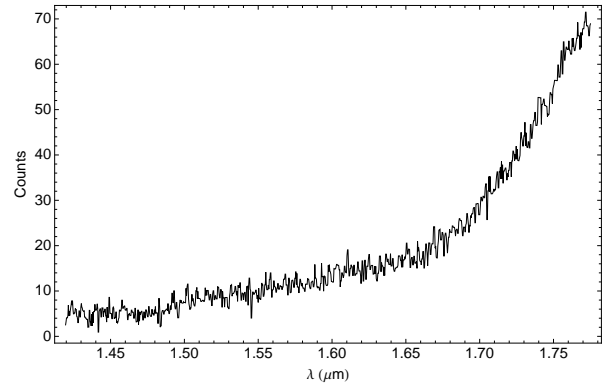} 
\caption{Spectrum in the central fiber of a 30 minute cold frame. This is used to remove the thermal background and detector noise from our blank sky observations.\label{fig:cold}}
\end{figure}

\begin{figure*}
\center
\includegraphics[width=1\textwidth]{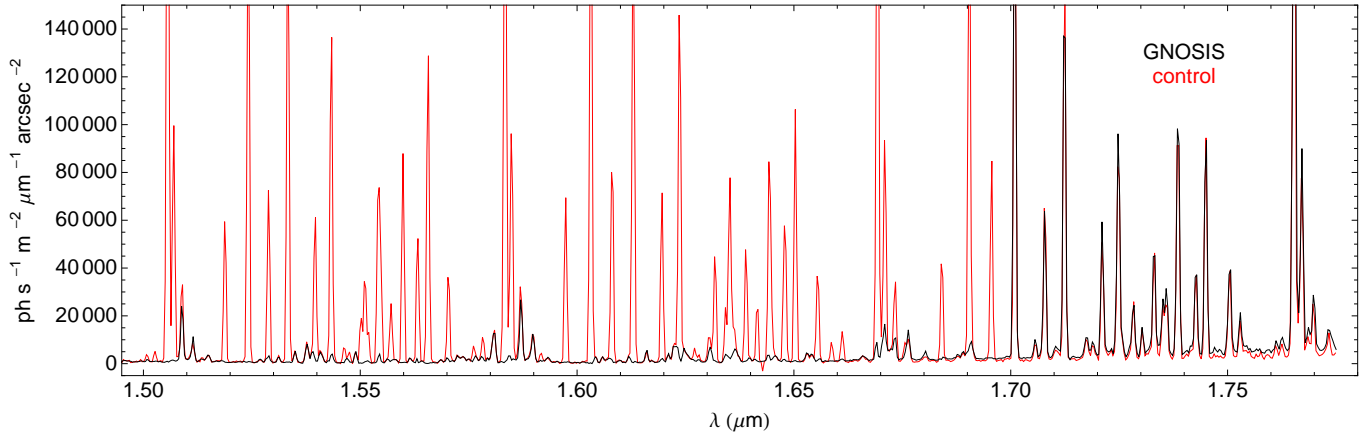} 
\caption{Spectrum of the night sky with (black) and without (red) OH suppression. The spectrum comes from 21 exposures (8.75\,hr total) from 1-4 September at various locations on the sky. The reduction in the OH lines in the range 1.5--1.7\,$\micron$ is clear.\label{fig:skyspec}}
\end{figure*}

\subsection{Sky Suppression}
The OH suppression performance of GNOSIS was measured from observations of blank sky. A total of 45 blank sky observations were taken during the September run at various locations on the sky. The exposure time of these observations was either 15 or 30 minutes. The instrument thermal background is significant in GNOSIS observations and a separate thermal background subtraction must be applied to blank sky observations. Cold frames, observations where the fore-optics unit was removed and pointed at a container of liquid nitrogen, were obtained with the same exposure time as the blank sky observations. Figure \ref{fig:cold} shows the smoothed cold frame spectrum from the central GNOSIS fiber. Alternatively, we may fit a thermal blackbody spectrum to the continuum points in between the OH lines $\lambda > 1.7\,\micron$ and subtract the best-fitting model from our spectrum.

Figure \ref{fig:skyspec} shows the cold-subtracted blank sky spectrum from six suppressed fibers and the control fiber from 8.75\,hr of observations from 2011 September 1 to 4. These 21 blank sky observations were made after moonset and are all $>$ 60$^{\circ}$ from the Moon. The spectra were flux-calibrated assuming an efficiency of 3.3\%. The strong suppression of the OH lines in the range 1.5--1.7\,$\micron$ is evident. 

\begin{figure}
\center
\includegraphics[width=0.45\textwidth]{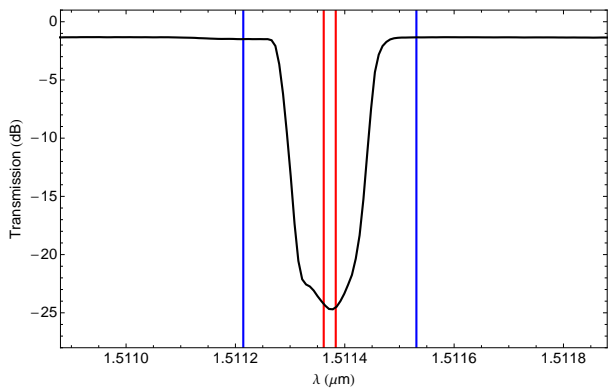} 
\caption{H1 FBG notch (black) designed to suppressed the 3--1 Q1 (4.5) OH doublet. The notch position was selected based on the mean wavelength of the doublet (red) according to \citet{rousselot2000}. The mean wavelength of the doublet (blue) according to \citet{abrams1994} is the same, but the spacing is much wider. This may be responsible for the strong residual seen at this wavelength in the OH-suppressed sky spectrum.\label{fig:abrams}}
\end{figure}

The suppression factor was measured for approximately half the lines and 78\% meet or exceed the target specifications. The other lines that do not meet the target specifications mostly correspond to FBG notches that are too narrow to suppress the entire doublet. The FBGs are designed based on the doublet separations found in the OH line model of \citet{rousselot2000}. The notches that do not meet the target specification have a much wider doublet separation according to \citet{abrams1994}. The doublet mean is the same, but the FBG notch is not wide enough to suppress the doublet according to \citet{abrams1994}. An example is shown in Figure \ref{fig:abrams} for the 3--1 Q1(4.5) transition, which shows the \citet{rousselot2000} values in red and the \citet{abrams1994} values in blue. 

Nevertheless, this is a clear demonstration that FBGs can cleanly remove OH lines while maintaining relatively high throughput between the lines. The integrated background in the range 1.5--1.7\,$\micron$ was reduced by a factor of $\approx$\,8, but there was no significant reduction in the interline background as predicted by \citet{sce2008}. This is unexpected given that \citet{ss2012} recently found that scattered OH dominates the interline background in their $H$-band spectrum. 

The blank sky observations made with GNOSIS thus far are detector noise-dominated in the interline regions. The faintness of the interline background and low system throughput of our current configuration requires 30 minute exposures for an interline background of $\approx$\,10\,ADU per pixel, which corresponds to $\approx$\,45\,$e^{-}$ per pixel. Of this, approximately 27\,$e^{-}$ per pixel are from the detector dark current and the other 18\,$e^{-}$ per pixel are from the sky. Thus, the detector's 8\,$e^{-}$ per pixel read noise gives $\approx$\,17\% uncertainty in the interline background. Thus, it is possible that our OH suppression fibers are reducing the interline background, but we were unable to observe the reduction of the low interline signal among the detector noise. Alternatively, there may be no reduction because of physical sources, but it is impossible to tell without observations that are not detector noise-dominated.




\subsection{Sensitivity}
The sensitivity of GNOSIS was measured from an observation of a low surface brightness galaxy HIZOA J0836-43 ($H$-band surface brightness of 17.3\,mag\,arcsec$^{2}$). The galaxy is larger than the GNOSIS FOV and the unknown aperture losses were not an issue. The sky-subtracted spectrum was divided by the square root of the non-subtracted spectrum to obtain the signal to noise per pixel. The median signal to noise per pixel was $\approx$\,10 for a 30 minute exposure. The same analysis on the spectrum through the control fiber yields the same result, surprisingly indicating that there was no improvement in signal to noise per pixel when using GNOSIS OH suppression. This is because OH suppression mainly improves the signal to noise near the OH lines. In between the lines (the region that dominates the sensitivity calculation) the reduction in the background was much smaller than expected. 

\section{Discussion}\label{section:discussion}
In this paper, we have given a thorough description of each subsystem of GNOSIS, the first instrument to use fiber Bragg grating OH suppression fibers. These fibers were designed to demonstrate a reduction of the interline background increasing sensitivity by a factor of 40 or more when suppressing the brightest OH doublets in the $H$-band. The line spread function of IRIS2 shows that the spectrograph's diffraction grating scatters light many pixels from the line center and simulations show that the cumulative contribution from multiple lines leads to a continuum of scattered light that dominates the interline background \citep{sce2008}. If the emission lines are suppressed before the light has the opportunity to be scattered by the spectrograph, the continuum of scattered light and therefore the interline background will be reduced.

\begin{figure}[!t]
\center
\includegraphics[width=0.45\textwidth]{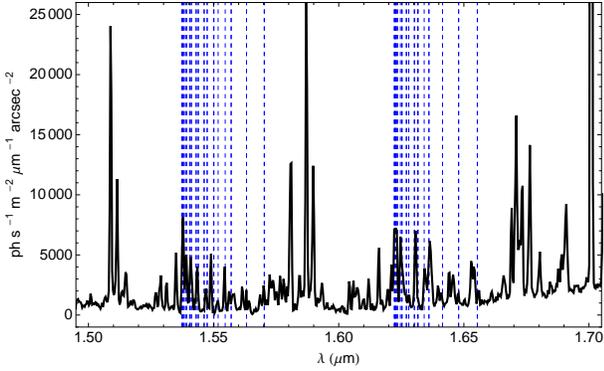} 
\caption{Detailed look at the OH-suppressed spectrum shown in Figure \ref{fig:skyspec} with the OH lines corresponding to $R$ transitions from \citet{abrams1994} shown by the dashed blue lines. These $R$ lines were not suppressed by design based on their relative strengths in \citet{rousselot2000} but they match up well with the residual features in the spectrum.\label{fig:abramsR}}
\end{figure}

To demonstrate this effect, we carried out observations of blank sky with GNOSIS at the 3.9\,m AAT with the IRIS2 spectrograph. Although the OH suppression fibers successfully suppressed most of the brightest OH doublets in the range 1.5--1.7\,$\micron$ at the target level or greater and reduced the integrated background by a factor of $\approx$\,10, the spectra showed no significant reduction in the interline background and thus the same median signal to noise per pixel when compared to a control fiber without OH suppression. A reduction in the interline background and the associated increase in sensitivity would led to OH suppression fibers having  an impact on astronomy at the level of adaptive optics or greater. Therefore, it is critically important to identify why no reduction was observed in these experiments and determine if improvements can be made to realize the full benefit of OH suppression.


Only observations that are detector noise-dominated have been made with OH suppression fibers. If no interline background reduction is seen in OH-suppressed observations that are not detector noise-dominated it would indicate either OH suppression fibers do not suppress scattered light or the atmosphere is more complex than modeled by \citet{sce2008}. Specifically, the OH spectrum may be different than the model of \citet{rousselot2000} used to design the FBGs, which would result in poor suppression. We have presented evidence of this above and poorly suppressed lines such as these would scatter more light into the interline regions. Alternatively, there may be unaccounted for atmospheric continuum sources from other molecular species that creates a continuum floor. It may also be possible that the unsuppressed OH lines are not as weak as indicated by \citet{rousselot2000} and they scatter more light into the interline regions. As evidence, we found that the $R$ transitions found in \citet{abrams1994} match up well with residual features in our OH-suppressed spectrum of the sky as seen in Figure \ref{fig:abramsR}. These doublets according to \citet{abrams1994} are stronger than indicated by \citet{rousselot2000} and would contribute more scattered light to the interline region.

The evidence that inaccuracies in the FBG design because of inaccuracies in the OH line model of \citet{rousselot2000}, which may result in more scattered light in the interline region than expected, is certainly suggestive. However, this is based on the assumption that OH suppression fibers suppress scattered light, which has yet to be demonstrated. Thus, it is critically important that the next step be a clear demonstration of scattered light suppression with observations that are not detector noise-dominated. 

Such observations require that the interline signal is greater and/or the detector noise is lower. In general, this may be accomplished by increasing the brightness of the source, increasing the throughput of the system and/or utilizing a low-noise detector. The system throughput may be increased mostly significantly by optimizing the OH suppression fibers and/or the spectrograph. Presently,  without first demonstrating that OH suppression fibers suppress scattered light, fabricating optimized fibers would be difficult to justify. Thus, \citet{sce2012} suggested that pairing GNOSIS with a fiber-fed spectrograph with a high efficiency volume phase holographic (VPH) grating is the best way to go about improving system throughput. If a high-performance detector, a 1.7\,$\micron$ cutoff Hawaii-2RG for example, were included the spectrograph would simultaneously increase throughput and lower detector noise, which would be ideal for OH-suppressed observations of blank sky. The Australian Astronomical Observatory has plans to build such a spectrograph called PRAXIS for further testing of OH suppression fibers on-sky by early 2014 \citep{horton2012}.

However, for the purpose of demonstrating scattered light suppression, observations of blank sky are not ideal because the sky spectrum may be considerably more complex than the spectrum the FBGs are designed to suppress. Thus, the ideal source would be an OH line source in the laboratory, such as that created by \citet{abrams1994}. This would remove any complications from additional atmospheric sources, but the weaker suppressed OH lines are still present. Alternatively, scattered light suppression may also be demonstrated with a single emission line from a bright arc lamp (xenon has at least one line at 1.6733\,$\micron$ that coincides with one of the FBG notches). These observations only require GNOSIS and the IRIS2 spectrograph without any telescope, which is advantageous because they do not require building a new spectrograph.

Regardless of whether or not OH suppression fibers can be used to reduce the interline background and increase sensitivity, they provide real benefits for observations at low resolving powers (500$<R<$3000), which was shown by \citet{sce2012} using observations of [\ion{Fe}{2}] emission lines in Seyfert galaxies and CH$_{4}$ absorption in brown dwarfs. Previously, observations at these resolving powers have been too low to resolve out the OH lines, but that is no longer necessary with OH suppression fibers.

The performance of the OH suppression fibers is very good, but there is room for improvement especially the throughput. The biggest single loss in the OH suppression fibers comes from MM to SM conversion of the input photonic lantern. The beam feeding the delivery fiber of the input photonic lanterns is $\approx$\,$f$/5 and we argued in Section \ref{section:photoniclantern} that the OH suppression fibers must be fed at $>$ $f$/7 to avoid overfilling the MM port ($f$/5.8) of the 1$\times$19 photonic lanterns. Thus, the MM to SM conversion throughput will increase if we slow the beam feeding the OH suppression fiber from $f$/5 to $f$/7. For a back-to-back system with 1$\times$19 photonic lanterns, the total throughput increases by $\approx$\,0.5\,dB (11\%) when slowing the input beam from $f$/5 to $f$/7 \citep{noordegraaf2012}. However, doing so will decrease the fiber's FOV. By conservation of \'etendue, the product of the fiber FOV on-sky and the telescope diameter is proportional to the product of the fiber input NA and the fiber core diameter.  Therefore, it is not possible to underfill the photonic lantern MM port by reducing the core diameter or input NA without sacrificing the fiber's FOV.

As an alternative to slowing the input beam from $f$/5 to $f$/7 to avoid overfilling the MM port, we may increase the MM port NA to match the input $f$/5 beam. If we keep the core diameter fixed, this corresponds to increasing $N$, the number of SMFs in the photonic lantern. Based on Equation (\ref{equation:lanternNA}) an MM port NA of 0.12 (the $f$/5 beam arrives at the MM port at $f$/4.2 due to FRD) with $d=50\,\micron$ corresponds to $N=37$. Thus, a back-to-back system with 1$\times$37 photonic lanterns fed at f/5 would be equivalent to a system with 1$\times$19 photonic lanterns fed at $f$/7 and should yield a similar increase in throughput over the current configuration. 

The wavelength-dependent loss of a symmetric photonic lantern system may be addressed by designing the MM port of the input and output photonic lanterns to have different NAs. If $N$ is the number of SMFs then the MM port of the input photonic lantern should be designed to support $M(\mathrm{blue})=N$ modes at the blue end of the suppression range. As $M(\lambda)\propto\lambda^{-2}$ there will be $M(\mathrm{red}) < N$ modes at the red end of the suppression range and no penalty will be incurred in the MM to SM conversion at any wavelength. The MM port of the output photonic lantern should be designed to support $M(\mathrm{red})=N$ modes at the red end of the suppression range, which would correspond to $M(\mathrm{blue}) > N$ modes at the blue end of the suppression range. Thus, in the SM to MM conversion, none of the modes will be lost at any wavelength.
 




Photonic lanterns with large $N$ are very cumbersome to handle and the GNOSIS grating unit is rather bulky and heavy (100\,kg). A significant reduction in size and weight may be possible with OH suppression fibers consisting of FBGs printed in multi-core fibers (MCFs) with each end tapered down into an MMF. \citet{birks2012} demonstrated such a device with an MCF containing $\approx$\,120 cores within a 230\,$\micron$ cladding and a single notch. MCFs with 37 cores can be manufactured and it should be a simple matter to taper down each end of the MCF into a 50\,$\micron$ MMF following a similar process for the manufacturing of the photonic lanterns. In addition to the reduction in size and weight no splices would be required boosting the throughput by $\approx$\,0.35\,dB (8\%). Assuming complex refractive index modulations can be imprinted into each MCF core with the same level of performance as the current GNOSIS FBGs and incorporating all the suggestions above would result in a throughput of at least $\approx$\,0.73. These next-generation OH suppression fibers are currently under development \citep{min2012}. 

Thus far we have discussed improvements to the OH suppression fibers themselves but the other subsystems also require optimization. Retrofitting GNOSIS to the existing IRIS2 spectrograph was acceptable for an initial demonstration of OH suppression fibers, but a fiber-fed spectrograph like PRAXIS would be ideal for future science observations. In the current configuration, the instrument thermal background is high (see Figure \ref{fig:cold}) and emanates almost entirely from the slit block of the IRIS2 interface unit reducing the sensitivity of observations. A spectrograph with a vacuum feed-through would make the IRIS2 interface unit unnecessary and significantly reduce the thermal background. Also, it would reduce the number of optical surfaces and there would be no slit block alignment errors, which would increase the system throughput by $\approx$\,15\% in addition to the increase in throughput due to the high efficiency VPH grating and the increase in sensitivity due to the high performance detector. 

OH suppression fibers have the potential to significantly expand our window to the universe from the ground if their suppression of scattered light can be confirmed and the atmosphere does not contain bright continuum sources that are un-filterable. Nevertheless, these fibers provide real benefits for spectroscopic observations at low resolving powers and optimized systems utilizing these types of fibers are very feasible.

\acknowledgments

The GNOSIS team acknowledges funding by ARC LIEF grant LE100100164. We thank the referee for providing insightful comments and suggestions which have greatly improved this paper. C.Q.T. gratefully acknowledges support by the National Science Foundation Graduate Research Fellowship under grant No. DGE-1035963. C.Q.T. thanks Stuart Gilchrist, Stuart Jackson, Ren\'ee Pelton, Billy Robbins, Andrew Sheinis, and Tim White.

\end{document}